\documentclass[sigconf]{acmart}

\makeatletter
\renewcommand\@formatdoi[1]{\ignorespaces}
\makeatother

\renewcommand\footnotetextcopyrightpermission[1]{} 
\acmDOI{}

\newcommand{\ignore}[1]{}
\usepackage{booktabs} 
\usepackage{array}
\renewcommand{\arraystretch}{1.2}  
\usepackage[normalem]{ulem}
\usepackage{hyperref}
\usepackage{subfiles}               
\usepackage{csvsimple}              
\usepackage{caption}                
\usepackage{subcaption}             
\usepackage{float}                  
\usepackage{amsfonts}               
\usepackage{amsmath}
\usepackage{graphicx}               
\usepackage{epstopdf}               
\usepackage{pbox}                   
\usepackage[boxed]{algorithm2e}     
\usepackage{multirow}               
\usepackage{threeparttable}         
\usepackage{xargs}                  
\usepackage{MnSymbol}               
\usepackage[binary-units=true]{siunitx}
\usepackage{capt-of}

\usepackage[colorinlistoftodos,prependcaption,textsize=small,disable]{todonotes}
\usepackage{marginnote}
\let\marginpar\marginnote
\newcommandx{\add}[2][1=]{\todo[inline,linecolor=blue,backgroundcolor=blue!25,bordercolor=blue,#1]{#2}}
\newcommandx{\unsure}[2][1=]{\todo[inline,linecolor=orange,backgroundcolor=orange!25,bordercolor=orange,#1]{#2}}
\newcommandx{\change}[2][1=]{\todo[inline,linecolor=red,backgroundcolor=red!25,bordercolor=red,#1]{#2}}
\newcommandx{\info}[2][1=]{\todo[inline,linecolor=olive,backgroundcolor=olive!25,bordercolor=olive,#1]{#2}}
\newcommandx{\improvement}[2][1=]{\todo[inline,linecolor=violet,backgroundcolor=violet!25,bordercolor=violet,#1]{#2}}
\newcommandx{\thiswillnotshow}[2][1=]{\todo[disable,#1]{#2}}

\begin{document}
\title[Dwarfs on Accelerators]{Dwarfs on Accelerators: Enhancing OpenCL Benchmarking for Heterogeneous Computing Architectures}

\author{Beau Johnston}
\orcid{0000-0001-5426-1415}
\affiliation{%
	\institution{Research School of Computer Science\\Australian National University}
	\city{Canberra} 
	\state{Australia}
}
\email{beau.johnston@anu.edu.au}

\author{Josh Milthorpe}
\orcid{0000-0002-3588-9896}
\affiliation{%
	\institution{Research School of Computer Science\\Australian National University}
	\city{Canberra} 
	\state{Australia}
}
\email{josh.milthorpe@anu.edu.au}

\begin{abstract}

For reasons of both performance and energy efficiency, high performance computing (HPC) hardware is becoming increasingly heterogeneous.
The OpenCL framework supports portable programming across a wide range of computing devices and is gaining influence in programming next-generation accelerators.
To characterize the performance of these devices across a range of applications requires a diverse, portable and configurable benchmark suite, and OpenCL is an attractive programming model for this purpose.


We present an extended and enhanced version of the OpenDwarfs OpenCL benchmark suite, with a strong focus placed on the robustness of applications, curation of additional benchmarks with an increased emphasis on correctness of results and choice of problem size.
Preliminary results and analysis are reported for eight benchmark codes on a diverse set of architectures -- three Intel CPUs, five Nvidia GPUs, six AMD GPUs and a Xeon Phi.




\end{abstract}


\maketitle

\section{Introduction}\label{sec:introduction}
	
High performance computing (HPC) hardware is becoming increasingly heterogeneous.
A major motivation for this is to reduce energy use; indeed, without significant improvements in energy efficiency, the cost of exascale computing will be prohibitive.
From June 2016 to June 2017, the average energy efficiency of the top 10 of the Green500 supercomputers rose by 2.3x, from 4.8 to 11.1 gigaflops per watt.\cite{feldman_2017}
For many systems, this was made possible by highly energy-efficient Nvidia Tesla P100 GPUs.
In addition to GPUs, future HPC architectures are also likely to include nodes with FPGA, DSP, ASIC and MIC components.
A single node may be heterogeneous, containing multiple different computing devices; moreover, a HPC system may offer nodes of different types.
For example, the Cori system at Lawrence Berkeley National Laboratory comprises 2,388 Cray XC40 nodes with Intel Haswell CPUs, and 9,688 Intel Xeon Phi nodes~\cite{declerck2016cori}.
The Summit supercomputer at Oak Ridge National Laboratory is based on the IBM Power9 CPU, which includes both NVLINK~\cite{morgan_2016}, a high bandwidth interconnect between Nvidia GPUs; and CAPI, an interconnect to support FPGAs and other accelerators.~\cite{morgan_2017}
Promising next generation architectures include Fujitsu's Post-K~\cite{morgan_2016_postk}, and Cray's CS-400, which forms the platform for the Isambard supercomputer~\cite{feldman_2017_isambard}.
Both architectures use ARM cores alongside other conventional accelerators, with several Intel Xeon Phi and Nvidia P100 GPUs per node.

Given this heterogeneity of hardware and the wide diversity of scientific application codes, workload characterization, performance prediction and scheduling are all becoming more challenging.
To evaluate different approaches requires a representative benchmark suite which is portable to a wide variety of devices.
We focus on the OpenCL programming model as it is supported on a wide range of systems including CPU, GPU and FPGA devices.
While it is possible to write application code directly in OpenCL, it may also be used as a base to implement higher-level programming models.
This technique was shown by \citet{mitra2014implementation} where an OpenMP runtime was implemented over an OpenCL framework for Texas Instruments Keystone II DSP architecture.
Having a common back-end in the form of OpenCL allows a direct comparison of identical code across diverse architectures.

In this paper, we present an extended version of the OpenDwarfs benchmark suite, a set of OpenCL benchmarks for heterogeneous computing platforms.\cite{krommydas2016opendwarfs}
We added new benchmarks to improve the diversity of the suite, and made a number of modifications aimed at improving the reproducibility and interpretability of results, portability between devices, and flexibility of configuration including problem sizes.
We report preliminary results for a subset of the enhanced OpenDwarfs suite on a range of platforms including CPU, GPU and MIC devices.


\section{Enhancing the OpenDwarfs Benchmark Suite}\label{sec:extending_the_opendwarfs_benchmark_suite}

The OpenDwarfs benchmark suite comprises a variety of OpenCL codes, classified according to patterns of computation and communication known as the 13 Berkeley Dwarfs.\cite{asanovic2006landscape}
The original suite focused on collecting representative benchmarks for scientific applications, with a thorough diversity analysis to justify the addition of each benchmark to the corresponding suite.
We aim to extend these efforts to achieve a full representation of each dwarf, both by integrating other benchmark suites and adding custom kernels.

\citet{marjanovic2016hpc} argue that the selection of problem size for HPC benchmarking critically affects which hardware properties are relevant.
We have observed this to be true across a wide range of accelerators, therefore we have enhanced the OpenDwarfs benchmark suite to support running different problem sizes for each benchmark.
To improve reproducibility of results, we also modified each benchmark to execute in a loop for a minimum of two seconds, to ensure that sampling of execution time and performance counters was not significantly affected by operating system noise.

For the Spectral Methods dwarf, the original OpenDwarfs version of the FFT benchmark was complex, with several code paths that were not executed for the default problem size, and returned incorrect results or failures on some combinations of platforms and problem sizes we tested.
We replaced it with a simpler high-performance FFT benchmark created by Eric Bainville~\cite{bainville2010fft}, which worked correctly in all our tests.
We have also added a 2-D discrete wavelet transform from the Rodinia suite~\cite{che2009rodinia} (with modifications to improve portability), and we plan to add a continuous wavelet transform code.

To understand benchmark performance, it is useful to be able to collect hardware performance counters associated with each timing segment.
LibSciBench is a performance measurement tool which allows high precision timing events to be collected for statistical analysis~\cite{hoefler2015scientific}.
It offers a high resolution timer in order to measure short running kernel codes, reported with one cycle resolution and roughly \SI{6}{\nano\second} of overhead.
We used LibSciBench to record timings in conjunction with hardware events, which it collects via PAPI~\cite{mucci1999papi} counters.
We modified the applications in the OpenDwarfs benchmark suite to insert library calls to LibSciBench to record timings and PAPI events for the three main components of application time: kernel execution, host setup and memory transfer operations.
Through PAPI modules such as Intel's Running Average Power Limit (RAPL) and Nvidia Management Library (NVML), LibSciBench also supports energy measurements, for which we report preliminary results in this paper.

\section{Related Work}\label{sec:related_work}
	
The NAS parallel benchmarks~\cite{bailey1991parallel} follow a `pencil-and-paper' approach, specifying the computational problem but leaving implementation choices such as language, data structures and algorithms to the user.
The benchmarks include varied kernels and applications which allow a nuanced evaluation of a complete HPC system, however, the unconstrained approach does not readily support direct performance comparison between different hardware accelerators using a single set of codes.

Martineau et al.~\cite{martineau2016performance} collected a suite of benchmarks and three mini-apps to evaluate Clang OpenMP 4.5 support for Nvidia GPUs.
Their focus was on comparison with CUDA; OpenCL was not considered.

The Scalable Heterogeneous Computing benchmark suite (SHOC)~\cite{lopez2015examining}, unlike OpenDwarfs and Rodinia, supports multiple nodes using MPI for distributed parallelism.
SHOC supports multiple programming models including OpenCL, CUDA and OpenACC, with benchmarks ranging from targeted tests of particular low-level hardware features to a handful of application kernels.
Sun et al.~\cite{sun2016} propose Hetero-Mark, a Benchmark Suite for CPU-GPU Collaborative Computing, which has five benchmark applications each implemented in HCC -- which compiles to OpenCL, HIP -- for a CUDA and Radeon Open Compute back-end, and a CUDA version.
Meanwhile, Chai by G{\'o}mez-Luna et al.~\cite{gomez2017chai}, offers 15 applications in 7 different implementations with the focus on supporting integrated architectures.

These benchmark suites focus on comparison between languages and environments; whereas our work focuses on benchmarking for device specific performance limitations, for example, by examining the problem sizes where these limitations occur -- this is largely ignored by benchmarking suites with fixed problem sizes.

Additionally, our enhanced OpenDwarfs benchmark suite aims to cover a wider range of application patterns by focusing exclusively on OpenCL using higher-level benchmarks.

\citet{barnes2016evaluating} collected a representative set of applications from the current NERSC workload to guide optimization for Knights Landing in the Cori supercomputer.
As it is not always feasible to perform such a detailed performance study of the capabilities of different computational devices for particular applications, the benchmarks described in this paper may give a rough understanding of device performance and limitations.

\section{Experimental Setup}\label{sec:experimental_setup}
\subsection{Hardware}\label{ssec:hardware}
\begin{table*}[t]
\caption{Hardware}
\centering
\begin{threeparttable}
    \centering
    \begin{tabular}{l|c|c|c|r|c|c|r|c}
        Name         & Vendor   & Type  & Series    & \multicolumn{1}{m{1cm}|}{\centering Core Count} & \multicolumn{1}{m{2.5cm}|}{\centering Clock Frequency (\si{\mega\hertz}) (min/max/turbo)}  &\multicolumn{1}{m{2.1cm}|}{\centering Cache (\SI{}{\kibi\byte}) (L1/L2/L3)} & \multicolumn{1}{m{.8cm}|}{\centering TDP (\SI{}{\watt})} &  \multicolumn{1}{m{1cm}}{\centering Launch  Date} \\ \hline
        Xeon E5-2697 v2  & Intel    & CPU   &Ivy Bridge & 24$\ast$ &1200/2700/3500 & 32/256/30720 & 130 & Q3 2013\\
        i7-6700K & Intel    & CPU   &Skylake & 8$\ast$ & 800/4000/4300 & 32/256/8192& 91 & Q3 2015\\
        i5-3550  & Intel    & CPU   & Ivy Bridge & 4$\ast$ & 1600/3380/3700 & 32/256/6144& 77 & Q2 2012\\
        Titan X & Nvidia & GPU & Pascal & 3584\textdagger & 1417/1531/-- & 48/2048/-- & 250 & Q3 2016\\
        GTX 1080 & Nvidia & GPU & Pascal & 2560\textdagger & 1607/1733/-- & 48/2048/-- & 180 & Q2 2016\\
        GTX 1080 Ti & Nvidia & GPU & Pascal & 3584\textdagger & 1480/1582/-- & 48/2048/-- & 250 & Q1 2017\\
        K20m & Nvidia & GPU & Kepler & 2496\textdagger & 706/--/-- & 64/1536/-- & 225 & Q4 2012\\
        K40m & Nvidia & GPU & Kepler & 2880\textdagger & 745/875/-- & 64/1536/-- & 235 & Q4 2013\\
        FirePro S9150 & AMD & GPU & Hawaii & 2816$\|$ & 900/--/-- & 16/1024/-- & 235 & Q3 2014\\
        HD 7970       & AMD & GPU & Tahiti & 2048$\|$ & 925/1010/-- & 16/768/-- & 250 & Q4 2011\\
        R9 290X       & AMD & GPU & Hawaii & 2816$\|$ & 1000/--/-- & 16/1024/--& 250 & Q3 2014\\
        R9 295x2      & AMD & GPU & Hawaii & 5632$\|$ & 1018/--/-- & 16/1024/--& 500 & Q2 2014\\
        R9 Fury X     & AMD & GPU & Fuji   & 4096$\|$ & 1050/--/-- & 16/2048/--& 273 & Q2 2015\\
        RX 480        & AMD & GPU & Polaris& 4096$\|$ & 1120/1266/-- & 16/2048/-- & 150 & Q2 2016\\
        Xeon Phi 7210 & Intel & MIC & KNL & 256\textdaggerdbl & 1300/1500/-- & 32/1024/-- & 215 & Q2 2016\\
    \end{tabular}
    \begin{tablenotes}
    \item [$\ast$] HyperThreaded cores
    \item [\textdagger] CUDA cores
    \item [$\|$] Stream processors
    \item [\textdaggerdbl] Each physical core has 4 hardware threads per core, thus 64 cores
    \end{tablenotes}
\end{threeparttable}
\label{tab:hardware}
\end{table*}

The experiments were conducted on a varied set of 15 hardware platforms: three Intel CPU architectures, five Nvidia GPUs, six AMD GPUs, and one MIC (Intel Knights Landing Xeon Phi).
Key characteristics of the test platforms are presented in Table~\ref{tab:hardware}.
The L1 cache size should be read as having both an instruction size cache and a data cache size of equal values as those displayed. 
For Nvidia GPUs, the L2 cache size reported is the size L2 cache per SM multiplied by the number of SMs.
For the Intel CPUs, hyper-threading was enabled and the frequency governor was set to `performance'.

\subsection{Software}\label{ssec:software}

OpenCL version 1.2 was used for all experiments.
On the CPUs we used the Intel OpenCL driver version 6.3, provided in 16.1.1 and the 2016-R3 opencl-sdk release.
On the Nvidia GPUs we used the Nvidia OpenCL driver version 375.66, provided as part of CUDA 8.0.61, AMD GPUs used the OpenCL driver version provided in the amdappsdk v3.0.

The Knights Landing (KNL) architecture used the same OpenCL driver as the Intel CPU platforms, however, the 2018-R1 release of the Intel compiler was required to compile for the architecture natively on the host.
Additionally, due to Intel removing support for OpenCL on the KNL architecture, some additional compiler flags were required.
Unfortunately, as Intel has removed support for AVX2 vectorization (using the `{\tt -xMIC-AVX512}' flag), vector instructions use only 256-bit registers instead of the wider 512-bit registers available on KNL.
This means that floating-point performance on KNL is limited to half the theoretical peak.

GCC version 5.4.0 with glibc 2.23 was used for the Skylake i7 and GTX 1080,  
GCC version 4.8.5 with glibc 2.23 was used on the remaining platforms.
OS Ubuntu Linux 16.04.4 with kernel version 4.4.0 was used for the Skylake CPU and GTX 1080 GPU, Red Hat 4.8.5-11 with kernel version 3.10.0 was used on the other platforms.

As OpenDwarfs has no stable release version, it was extended from the last commit by the maintainer on 26 Feb 2016.~\cite{opendwarfs2017base}
LibSciBench version 0.2.2 was used for all performance measurements.

\subsection{Measurements}\label{ssec:measurements}


We measured execution time and energy for individual OpenCL kernels within each benchmark.
Each benchmark run executed the application in a loop until at least two seconds had elapsed, and the mean execution time for each kernel was recorded.
Each benchmark was run 50 times for each problem size (see \S\ref{ssec:setting_sizes}) for both execution time and energy measurements.
A sample size of 50 per group -- for each combination of benchmark and problem size -- was used to ensure that sufficient statistical power $\beta = 0.8$ would be available to detect a significant difference in means on the scale of half standard deviation of separation.
This sample size was computed using the t-test power calculation over a normal distribution.

To help understand the timings, the following hardware counters were also collected:
\begin{itemize}
	\item total instructions and IPC (Instructions Per Cycle);
	\item L1 and L2 data cache misses;
	\item total L3 cache events in the form of request rate (requests / instructions), miss rate (misses / instructions), and miss ratio (misses/requests);
	\item data TLB (Translation Look-aside Buffer) miss rate (misses / instructions); and
	\item branch instructions and branch mispredictions.
\end{itemize}
For each benchmark we also measured memory transfer times between host and device, however, only the kernel execution times and energies are presented here.

Energy measurements were taken on Intel platforms using the RAPL PAPI module, and on Nvidia GPUs using the NVML PAPI module.

\subsection{Setting Sizes}\label{ssec:setting_sizes}

For each benchmark, four different problem sizes were selected, namely {\bf tiny}, {\bf small}, {\bf medium} and {\bf large}.
These problem sizes are based on the memory hierarchy of the Skylake CPU.
Specifically, {\bf tiny} should just fit within L1 cache, on the Skylake this corresponds to \SI{32}{\kibi\byte} of data cache, {\bf small} should fit within the \SI{256}{\kibi\byte} L2 data cache, {\bf medium} should fit within \SI{8192}{\kibi\byte} of the L3 cache, and {\bf large} must be much larger than \SI{8192}{\kibi\byte} to avoid caching and operate out of main memory.

The memory footprint was verified for each benchmark by printing the sum of the size of all memory allocated on the device.

For this study, problem sizes were not customized to the memory hierarchy of each platform, since the CPU is the most sensitive to cache performance.
Also, note for these CPU systems the L1 and L2 cache sizes are identical, and since we ensure that {\bf large} is at least $4\times$ larger than L3 cache, we are guaranteed to have last-level cache misses for the {\bf large} problem size.

Caching performance was measured using PAPI counters.
On the Skylake L1 and L2 data cache miss rates were counted using the {\tt PAPI\_L1\_DCM} and {\tt PAPI\_L2\_DCM} counters.
For L3 miss events, only the total cache counter event ({\tt PAPI\_L3\_TCM}) was available.
The final values presented as miss results are presented as a percentage, and were determined using the number of misses counted divided by the total instructions ({\tt PAPI\_TOT\_INS}).

The methodology to determine the appropriate size parameters is demonstrated on the k-means benchmark.

\subsubsection{kmeans}
K-means is an iterative algorithm which groups a set of points into clusters, such that each point is closer to the centroid of its assigned cluster than to the centroid of any other cluster.
Each step of the algorithm assigns each point to the cluster with the closest centroid, then relocates each cluster centroid to the mean of all points within the cluster.
Execution terminates when no clusters change size between iterations.
Starting positions for the centroids are determined randomly.
The OpenDwarfs benchmark previously required the object features to be read from a previously generated file.
We extended the benchmark to support generation of a random distribution of points.
This was done to more fairly evaluate cache performance, since repeated runs of clustering on the same feature space (loaded from file) would deterministically generate similar caching behavior.
For all problem sizes, the number of clusters is fixed at 5.

Given a fixed number of clusters, the parameters that may be used to select a problem size are the number of points $P_n$, and the dimensionality or number of features per point $F_n$.
In the kernel for k-means there are three large one-dimensional arrays passed to the device, namely {\bf feature}, {\bf cluster} and {\bf membership}.
In the {\bf feature} array which stores the unclustered feature space, each feature is represented by a 32-bit floating-point number, so the entire array is of size $P_n \times F_n \times \text{sizeof}\left(\text{float}\right)$.
{\bf cluster} is the working and output array to store the intermediately clustered points, it is of size $C_n \times F_n \times \text{sizeof}\left(\text{float}\right)$, where $C_n$ is the number of clusters.
{\bf membership} is an array indicating whether each point has changed to a new cluster in each iteration of the algorithm, it is of size $P_n \times \text{sizeof}\left(\text{int}\right)$, where $\text{sizeof}\left(\text{int}\right)$ is the number of bytes to represent an integer value.
Thereby the working kernel memory, in \si{\kibi\byte}, is:
\begin{equation}
    \frac{\text{size}\left(\textbf{feature}\right)+\text{size}\left(\textbf{membership}\right)+\text{size}\left(\textbf{cluster}\right)}{1024}
    \label{eq:kmeans_size}
\end{equation}

Using this equation, we can determine the largest problem size that will fit in each level of cache.
The tiny problem size is defined to have 256 points and 30 features; from Equation~\ref{eq:kmeans_size} the total size of the main arrays is \SI{31.5}{\kibi\byte}, slightly smaller than the \SI{32}{\kibi\byte} L1 cache.
The number of points is increased for each larger problem size to ensure that the main arrays fit within the lower levels of the cache hierarchy, measuring the total execution time and respective caching events.
The {\bf tiny}, {\bf small} and {\bf medium} problem sizes in the first row of Table~\ref{tab:problem_sizes} correspond to L1, L2 and L3 cache respectively.
The {\bf large} problem size is at least four times the size of the last-level cache -- in the case of the Skylake, at least \SI{32}{\mebi\byte} -- to ensure that data are transferred between main memory and cache.


For brevity, cache miss results are not presented in this paper but were used to verify the selection of suitable problem sizes for each benchmark.
The procedure to select problem size parameters is specific to each benchmark, but follows a similar approach to k-means.

\subsubsection{lud, fft, srad, crc, nw}
The LU-Decomposition {\tt lud}, Fast Fourier Transform {\tt fft}, Speckle Reducing Anisotropic Diffusion {\tt srad}, Cyclic Redundancy Check {\tt crc} and Needleman-Wunsch {\tt nw} benchmarks did not require additional data sets.
Where necessary these benchmarks were modified to generate the correct solution and run on modern architectures.
Correctness was examined either by directly comparing outputs against a serial implementation of the codes (where one was available), or by adding utilities to compare norms between the experimental outputs.

\subsubsection{dwt}
Two-Dimensional Discrete Wavelet Transform is commonly used in image compression.
It has been extended to support loading of Portable PixMap (.ppm) and Portable GrayMap (.pgm) image format, and storing Portable GrayMap images of the resulting DWT coefficients in a visual tiled fashion.
The input image dataset for various problem sizes was generated by using the resize capabilities of the ImageMagick application.
The original gum leaf image is the large sample size has the ratio of $3648 \times 2736$ pixels and was down-sampled to  $80 \times 60$.

\subsubsection{gem, nqueens, hmm}
For three of the benchmarks, we were unable to generate different problem sizes to properly exercise the memory hierarchy.

Gemnoui {\tt gem} is an n-body-method based benchmark which computes electrostatic potential of biomolecular structures.
Determining suitable problem sizes was performed by initially browsing the National Center for Biotechnology Information's Molecular Modeling Database (MMDB)~\cite{madej2013mmdb} and inspecting the corresponding Protein Data Bank format (pdb) files.
Molecules were then selected based on complexity, since the greater the complexity the greater the number of atoms required for the benchmark and thus the larger the memory footprint.
{\bf tiny} used the Prion Peptide 4TUT~\cite{yu2015crystal} and was the simplest structure, consisting of a single protein (1 molecule), it had the device side memory usage of \SI{31.3}{\kibi\byte} which should fit in the L1 cache (\SI{32}{\kibi\byte}) on the Skylake processor.
{\bf small} used a Leukocyte Receptor 2D3V~\cite{shiroishi2006crystal} also consisting of 1 protein molecule, with an associated memory footprint of 252KiB.
{\bf medium} used the nucleosome dataset originally provided in the OpenDwarfs benchmark suite, using \SI{7498}{\kibi\byte} of device-side memory.
{\bf large} used an X-Ray Structure of a Nucleosome Core Particle~\cite{davey2002solvent}, consisting of 8 protein, 2 nucleotide, and 18 chemical molecules, and requiring \SI{10970.2}{\kibi\byte} of memory when executed by {\tt gem}.
Each {\tt pdb} file was converted to the {\tt pqr} atomic particle charge and radius format using the {\tt pdb2pqr}~\cite{dolinsky2004pdb2pqr} tool.
Generation of the solvent excluded molecular surface used the tool {\tt msms}~\cite{sanner1996reduced}.
Unfortunately, the molecules used for the {\bf medium} and {\bf large} problem sizes contain uninitialized values only noticed on CPU architectures and as such further work is required to ensure correctness for multiple problem sizes.
The datasets used for {\tt gem} and all other benchmarks can be found in this paper's associated GitHub repository~\cite{johnston2017}.

The {\tt nqueens} benchmark is a backtrack/branch-and-bound code which finds valid placements of queens on a chessboard of size n$\times$n, where each queen cannot be attacked by another.
For this code, memory footprint scales very slowly with increasing number of queens, relative to the computational cost.
Thus it is significantly compute-bound and only one problem size is tested.

The Baum-Welch Algorithm Hidden Markov Model {\tt hmm} benchmark represents the Graphical Models dwarf and did not require additional data sets, however validation of the correctness of results has not occurred apart from over the {\bf tiny} problem size, as such, it is the only size examined in the evaluation.


\todo{Replace bfs if new results are available}

\subsubsection{Summary of Benchmark Settings}

The problem size parameters for all benchmarks are presented in Table~\ref{tab:problem_sizes}.

\begin{table}[thb]
	\centering
	\begin{threeparttable}
		\centering
		\caption{OpenDwarfs workload scale parameters $\Phi$}
		\begin{tabular}{l|c|c|c|c}
			\bf Benchmark         & \bf tiny   & \bf small  & \bf medium     & \bf large\\\hline
			{\tt kmeans}          & 256        & 2048   & 65600      & 131072\\
			{\tt lud}             & 80         & 240    & 1440       & 4096\\
			{\tt csr}             & 736        & 2416   & 14336      & 16384\\
			{\tt fft}             & 2048       & 16384  & 524288     & 2097152\\
			{\tt dwt}             & 72x54      & 200x150& 1152x864   & 3648x2736\\       
			{\tt srad}            & 80,16      & 128,80 & 1024,336   & 2048,1024\\
			{\tt crc}             & 2000       & 16000  & 524000     & 4194304\\
			\todo{Replace if new results are available}
            {\tt nw}              & 48         & 176    & 1008       & 4096\\
            {\tt gem}             & 4TUT       & 2D3V   & nucleosome & 1KX5\\
            {\tt nqueens}         & 18         & -- & -- & --\\
            {\tt hmm}             & 8,1        & 900,1  & 1012,1024  & 2048,2048\\
		\end{tabular}
		\label{tab:problem_sizes}
	\end{threeparttable}
\end{table}

Each {\bf Device} can be selected in a uniform way between applications using the same notation, on this system {\bf Device} comprises of {\tt -p 1 -d 0 -t 0} for the Intel Skylake CPU, where {\tt p} and {\tt d} are the integer identifier of the platform and device to respectively use, and {\tt -p 1 -d 0 -t 1} for the Nvidia GeForce GTX 1080 GPU.
Each application is run as {\bf Benchmark} {\bf Device} {\tt --} {\bf Arguments}, where {\bf Arguments} is taken from Table~\ref{tab:program_arguments} at the selected scale of $\Phi$.
For reproducibility the entire set of Python scripts with all problem sizes is available in a GitHub repository~\cite{johnston2017}. 
Where $\Phi$ is substituted as the argument for each benchmark, it is taken as the respective scale from Table~\ref{tab:problem_sizes} and is inserted into Table~\ref{tab:program_arguments}.

\begin{table}[t]
	\centering
	\begin{threeparttable}
		\centering
		\captionof{table}{Program Arguments}
		\vspace{0pt}
		\begin{tabular}{l|l}
			\bf Benchmark & \bf Arguments\\\hline
			{\tt kmeans} & {\tt -g -f 26 -p} $\Phi$\\
			{\tt lud} & {\tt -s} $\Phi$\\
			{\tt csr}\textdagger & {\tt -i} $\Psi$\\
			      & $\Psi$ = {\tt createcsr -n} $\Phi$ {\tt -d 5000} $\medtriangleup$\\
			{\tt fft} & $\Phi$ \\
			{\tt dwt} & {\tt -l 3 }$\Phi${\tt-gum.ppm}\\
			{\tt srad}& $\Phi_1$ $\Phi_2$ {\tt 0 127 0 127 0.5 1}\\
			{\tt crc}&  {\tt -i 1000\_}$\Phi${\tt.txt}\\
			\todo{Replace if new results are available}
            {\tt nw}&  $\Phi${ 10}\\
            {\tt gem} & $\Phi$ {80 1 0}\\
            {\tt n-queens} & $\Phi$\\
            {\tt hmm}&  {\tt -n }$\Phi_1${\tt -s }$\Phi_2${\tt -v s}\\
		\end{tabular}
		\begin{tablenotes}
			\item [$\medtriangleup$] The {\tt -d 5000} indicates density of the matrix in this instance 0.5\% dense (or 99.5\% sparse).
			\item [\textdagger] The {\tt csr} benchmark loads a file generated by {\tt createcsr} according to the workload size parameter $\Phi$; this file is represented by $\Psi$.
		\end{tablenotes}
		\label{tab:program_arguments}
	\end{threeparttable}
\end{table}

\section{Results}\label{sec:results}

The primary purpose of including these time results is to demonstrate the benefits of the extensions made to the OpenDwarfs Benchmark suite.
The use of LibSciBench allowed high resolution timing measurements over multiple code regions.
To demonstrate the portability of the Extended OpenDwarfs benchmark suite, we present results from 11 varied benchmarks running on 15 different devices representing four distinct classes of accelerator.
For 12 of the benchmarks, we measured multiple problem sizes and observed distinctly different scaling patterns between devices.
This underscores the importance of allowing a choice of problem size in a benchmarking suite.

\subsection{Time}\label{ssec:time}
	
We first present execution time measurements for each benchmark, starting with the Cyclic Redundancy Check {\tt crc} benchmark which represents the Combinational Logic dwarf.

\newcommand{\plotwidth}{0.24\textwidth}

\begin{figure*}[t]
	\centering
	\includegraphics[width=\textwidth,keepaspectratio]{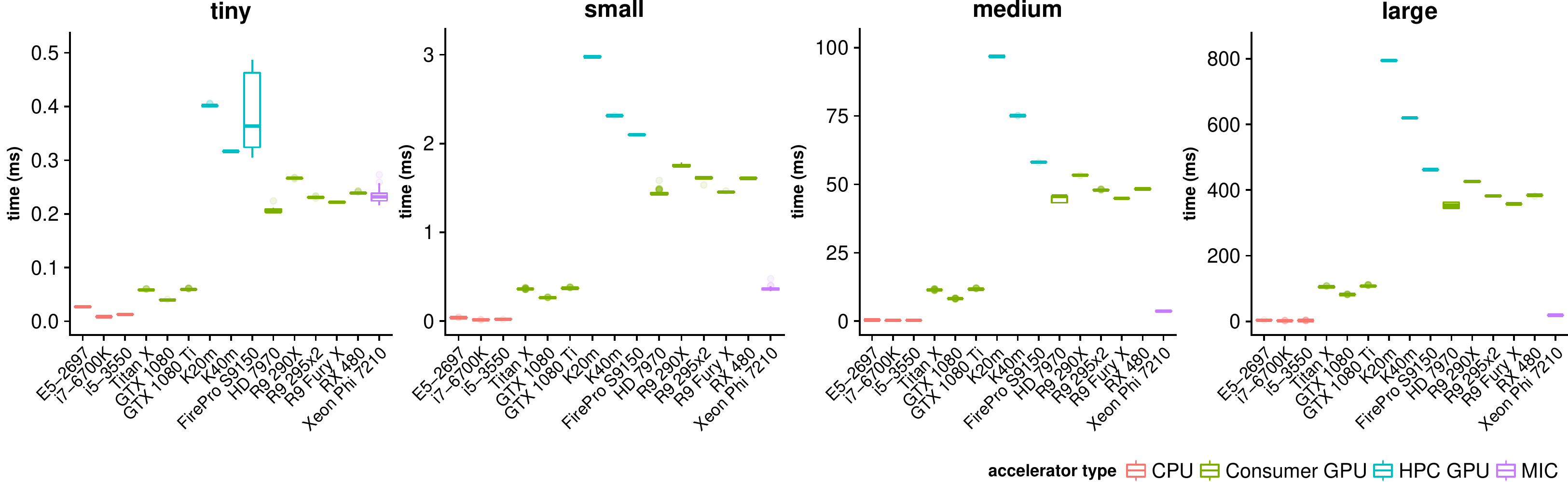}
	\caption{Kernel execution times for the {\bf crc} benchmark on different hardware platforms}
	\label{fig:time-crc}
\end{figure*}

Figure~\ref{fig:time-crc} shows the execution times for the {\tt crc} benchmark over 50 iterations on each of the target architectures, including the KNL.
The results are colored according to accelerator type: red for CPU devices, green for consumer GPUs, blue for HPC GPUs, and purple for the KNL MIC.
Execution times for {\tt crc} are lowest on CPU-type architectures, probably due to the low floating-point intensity of the CRC computation~\cite[Ch. 6]{joshi2016thesis}.
Excluding {\tt crc}, all the other benchmarks perform best on GPU type accelerators; furthermore, the performance on the KNL is poor due to the lack of support for wide vector registers in Intel's OpenCL SDK.
We therefore omit results for KNL for the remaining benchmarks.

\todo{We could examine the kiviat diagrams to see if they have high memory address entropy -- or high memory overhead which contributes to the why some benchmarks see a wide variation while others experience only a little, is this the major cause in variation?}
\todo{What are the major motivations for including this results section? We should conclude with these motivations and what has been shown accordingly}

\begin{figure*}
    \centering
    \includegraphics[width=.9\textwidth,keepaspectratio]{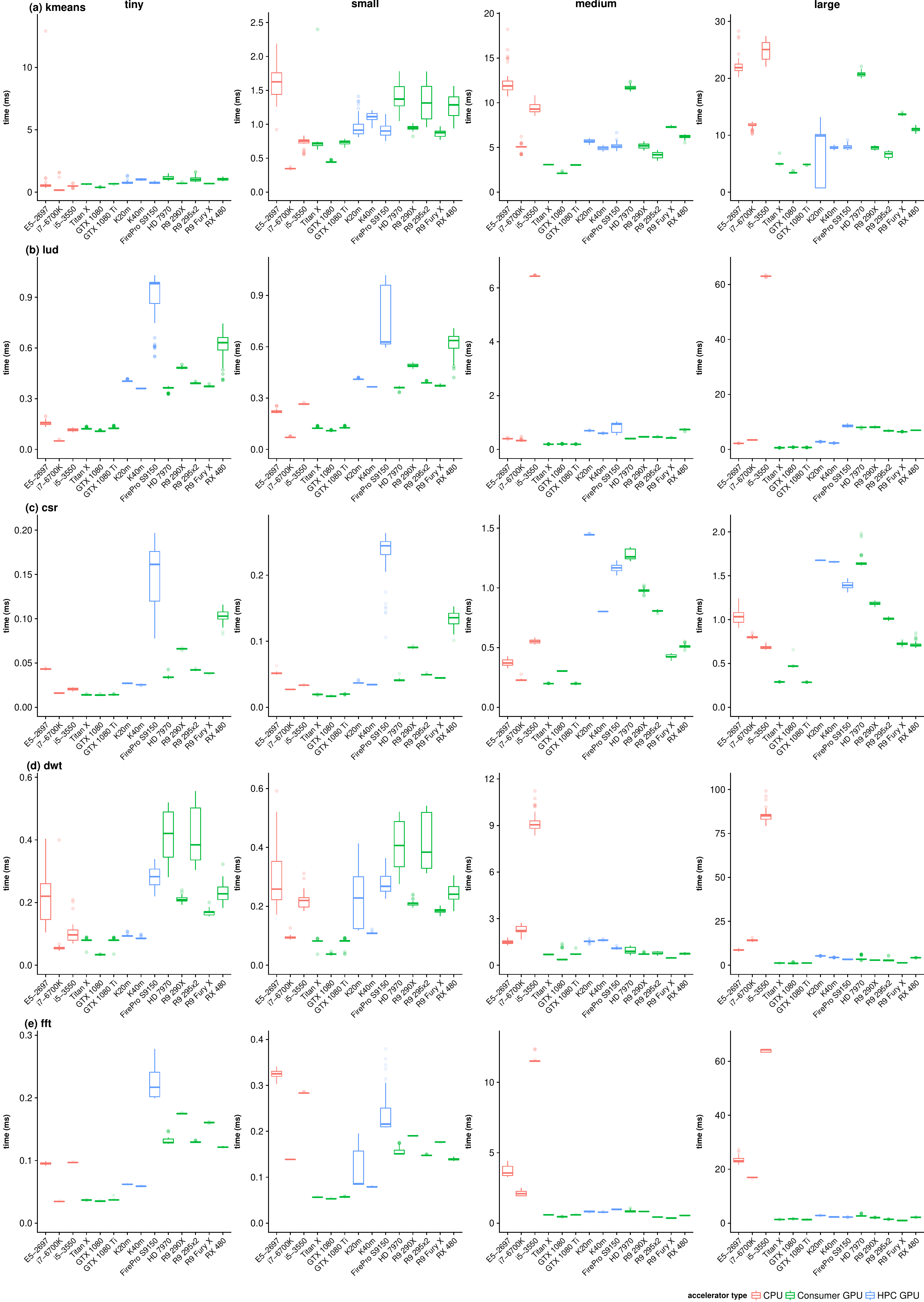}
    \caption{Benchmark kernel execution times on different hardware platforms}
    \label{fig:time}
\end{figure*}

\begin{figure*}[t]
    \centering
    \includegraphics[width=\textwidth,keepaspectratio]{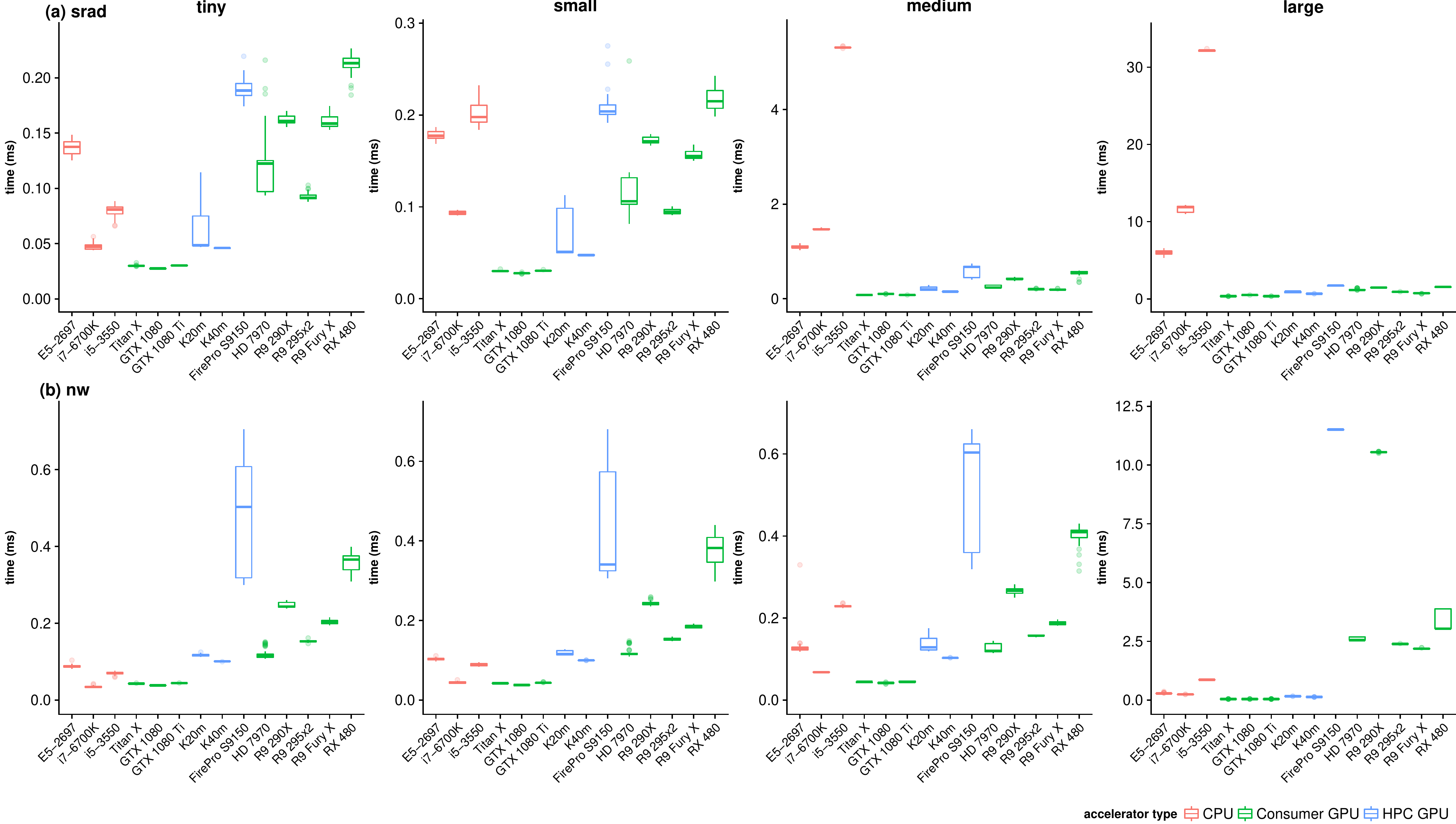}
    \caption{Benchmark kernel execution times on different hardware platforms (continued)}
    \label{fig:time2}
\end{figure*}

Figures~\ref{fig:time} and~\ref{fig:time2} shows the distribution of kernel execution times for the remaining benchmarks.
Some benchmarks execute more than one kernel on the accelerator device; the reported iteration time is the sum of all compute time spent on the accelerator for all kernels.
Each benchmark corresponds to a particular dwarf: 
Figure~\ref{fig:time}a ({\tt kmeans}) represents the MapReduce dwarf,
Figure~\ref{fig:time}b ({\tt lud}) represents the Dense Linear Algebra dwarf,
Figure~\ref{fig:time}c ({\tt csr}) represents Sparse Linear Algebra, 
Figure~\ref{fig:time}d ({\tt dwt}) and Figure~\ref{fig:time}e ({\tt fft}) represent Spectral Methods,
Figure~\ref{fig:time2}a ({\tt srad}) represents the Structured Grid dwarf and Figure~\ref{fig:time2}b ({\tt nw}) represents Dynamic Programming.

\begin{figure*}
    \centering
    \includegraphics[width=\textwidth,keepaspectratio]{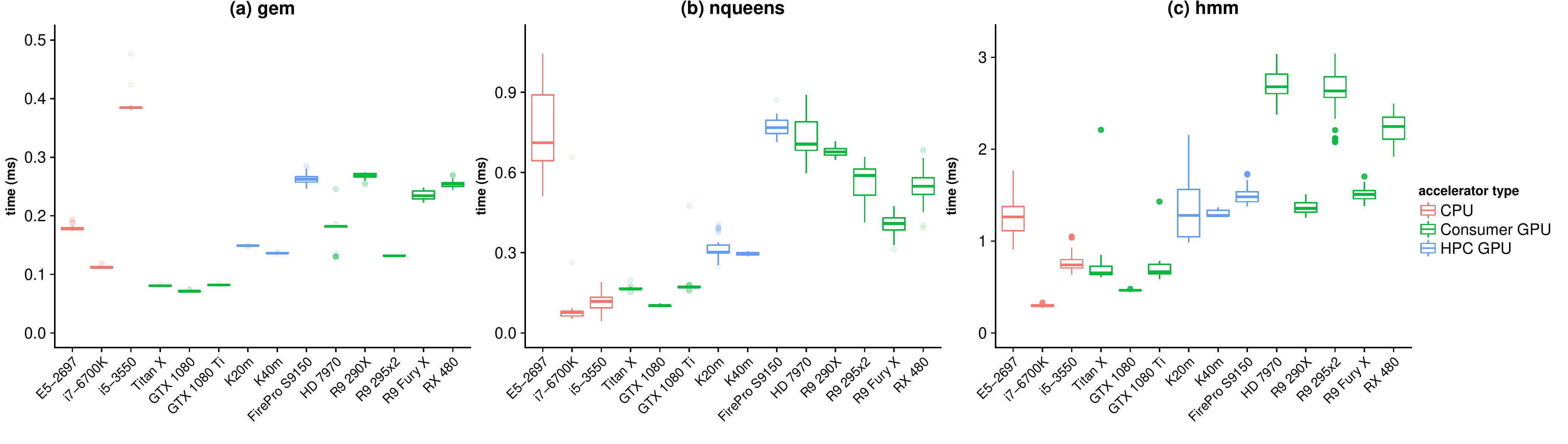}
    \caption{Single problem sized benchmarks of kernel execution times on different hardware platforms}
    \label{fig:time3}
\end{figure*}

Finally, Figure~\ref{fig:time3} presents results for the three applications with restricted problem sizes and only one problem size is shown.
The N-body Methods dwarf is represented by ({\tt gem}) and the results are shown in Figure~\ref{fig:time3}a, the Backtrack \& Branch and Bound dwarf is represented by the ({\tt nqueens}) application in Figure~\ref{fig:time3}b and ({\tt hmm}) results in Figure~\ref{fig:time3}c represent the Graphical Models dwarf.

Examining the transition from tiny to large problem sizes (from left to right) in Figure~\ref{fig:time2}a shows the performance gap between CPU and GPU architectures widening for {\tt srad} -- indicating codes representative of structured grid dwarfs are well suited to GPUs.

In contrast, Figure~\ref{fig:time2}b shows Dynamic Programming problems have performance results tied to micro-architecture or OpenCL runtime support and can not be explained solely by considering accelerator type.
For instance, the Intel CPUs and NVIDIA GPUs perform comparably over all problem sizes, whereas all AMD GPUs exhibit worse performance as size increases.

For most benchmarks, the coefficient of variation in execution times is much greater for devices with a lower clock frequency, regardless of accelerator type.
While execution time increases with problem size for all benchmarks and platforms, the modern GPUs (Titan X, GTX1080, GTX1080Ti, R9 Fury X and RX 480) performed relatively better for large problem sizes, possibly due to their greater second-level cache size compared to the other platforms.
A notable exception is {\tt k-means} for which CPU execution times were comparable to GPU, which reflects the relatively low ratio of floating-point to memory operations in the benchmark.

Generally, the HPC GPUs are older and were designed to alleviate global memory limitations amongst consumer GPUs of the time.
(Global memory size is not listed in Table~\ref{tab:hardware}.)
Despite their larger memory sizes, the clock speed of all HPC GPUs is slower than all evaluated consumer GPUs.
While the HPC GPUs (devices 7-9, in blue) outperformed consumer GPUs of the same generation (devices 10-13, in green) for most benchmarks and problem sizes, they were always beaten by more modern GPUs.
This is no surprise since all selected problem sizes fit within the global memory of all devices.

A comparison between CPUs (devices 1-3, in red) indicates the importance of examining multiple problem sizes.
Medium-sized problems were designed to fit within the L3 cache of the i7-6700K system, and this conveniently also fits within the L3 cache of the Xeon E5-2697 v2.
However, the older i5-3550 CPU has a smaller L3 cache and exhibits worse performance when moving from small to medium problem sizes, and is shown in Figures~\ref{fig:time}b,~\ref{fig:time}d,~\ref{fig:time}e and ~\ref{fig:time2}a,

Increasing problem size also hinders the performance in certain circumstances for GPU devices.
For example, Figure~\ref{fig:time2}b shows a widening performance gap over each increase in problem size between AMD GPUs and the other devices.

Predicted application properties for the various Berkeley Dwarfs are evident in the measured runtime results.
For example, Asanovi\'{c} et al.~\cite{asanovic2006landscape} state that applications from the Spectral Methods dwarf is memory latency limited.
If we examine {\tt dwt} and {\tt fft} -- the applications which represent Spectral Methods -- in Figure~\ref{fig:time}d and Figure~\ref{fig:time}e respectively, we see that for medium problem sizes the execution times match the higher memory latency of the L3 cache of CPU devices relative to the GPU counterparts.
The trend only increases with problem size: the large size shows the CPU devices frequently accessing main memory while the GPUs' larger memory ensures a lower memory access latency.
\todo{What is the memory size for the GPUs, and what is the difference in latency?}
It is expected if had we extended this study to an even larger problem size that would not fit on GPU global memory, much higher performance penalties would be experienced over GPU devices, since the PCI-E interconnect has a higher latency than a memory access to main memory from the CPU systems.
As a further example, Asanovi\'{c} et al.~\cite{asanovic2006landscape} state that the Structured Grid dwarf is memory bandwidth limited.
The Structured Grid dwarf is represented by the {\tt srad} benchmark shown in Figure~\ref{fig:time2}a.
GPUs exhibit lower execution times than CPUs, which would be expected in a memory bandwidth-limited code as GPU devices offer higher bandwidth than a system interconnect.

\todo[inline]{The remaining applications could be examined in terms of dwarf}

\subsection{Energy}\label{ssec:energy}
	
In addition to execution time, we are interested in differences in energy consumption between devices and applications.
We measured the energy consumption of benchmark kernel execution on the Intel Skylake i7-6700k CPU and the Nvidia GTX1080 GPU, using PAPI modules for RAPL and NVML. 
These were the only devices examined since collection of PAPI energy measurements (with LibSciBench) requires superuser access, and these devices were the only accelerators available with this permission.
The distributions were collected by measuring solely the kernel execution over a distribution of 50 runs.
RAPL CPU energy measurements were collected over all cores in package 0 {\tt rapl:::PP0\_ENERGY:PACKAGE0}.
NVML GPU energy was collected using the power usage readings {\tt nvml:::GeForce\_GTX\_1080:power} for the device and presents the total power draw (+/-5 watts) for the entire card -- memory and chip.
Measurements results converted to energy \SI{}{\joule} \todo[inline]{power (\SI{}{\joule\per\second})} from their original resolution \SI{}{\nano\joule} and \SI{}{\milli\watt} on the CPU and GPU respectively.

From the time results presented in Section~\ref{ssec:time} we see the largest difference occurs between CPU and GPU type accelerators at the {\bf large} problem size.
Thus we expect that the {\bf large} problem size will also show the largest difference in energy.

\begin{figure*}[htb]
\begin{subfigure}{.49\textwidth}
\centering
\includegraphics[width=1\textwidth]{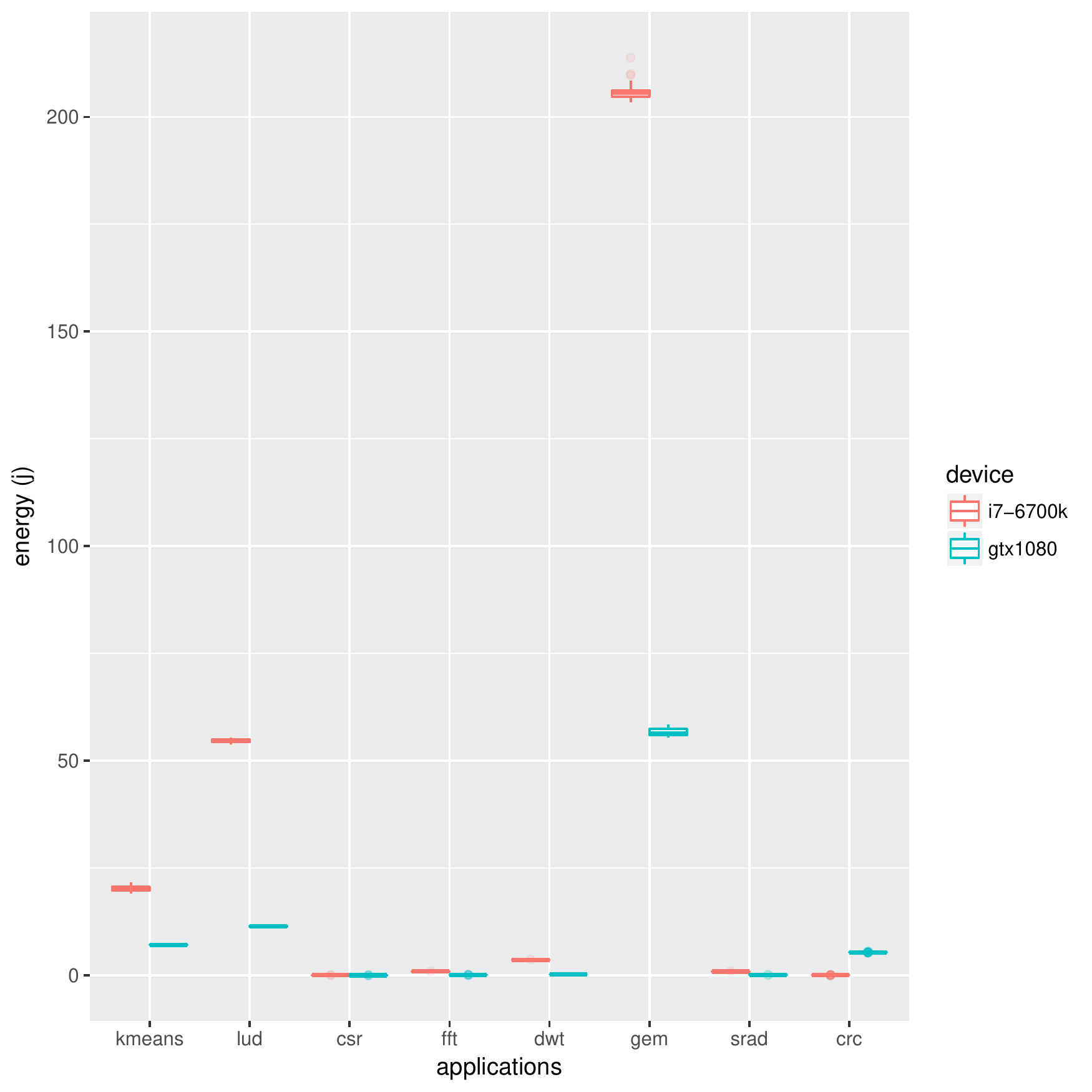}
\caption{Kernel execution energy}
\label{fig:energy}
\end{subfigure}
\hfill
\begin{subfigure}{.49\textwidth}
\centering
\includegraphics[width=1\textwidth]{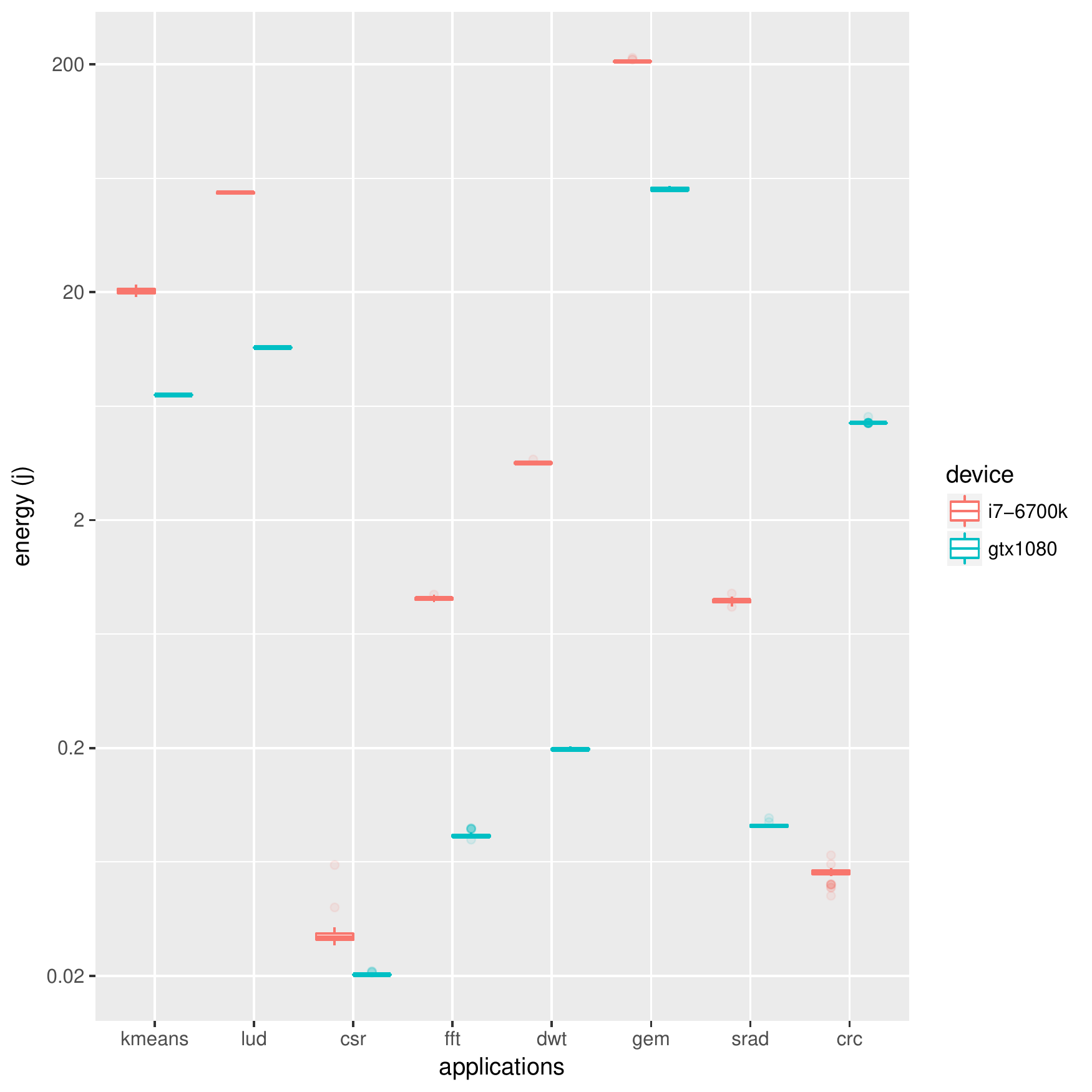}
\caption{Log of kernel execution energy}
\label{fig:energy-log}
\end{subfigure}
\caption{Benchmark kernel execution energy ({\bf large} problem size) on Core i7-6700K and Nvidia GTX1080}
\end{figure*}

\todo{Does energy scale with problem size for all benchmarks? Maybe not since dwarfs with large memory access cache misses could add huge overheads}

Figures~\ref{fig:energy} and~\ref{fig:energy-log} show the kernel execution energy for several benchmarks for the {\bf large} size.
All results are presented in joules.
The box plots are coloured according to device: red for the Intel Skylake i7-6700k CPU and blue for the Nvidia GTX1080 GPU.
\todo{report power as well as energy; proper analysis of the results}
The logarithmic transformation has been applied to Figure~\ref{fig:energy-log} to emphasise the variation at smaller energy scales ($<$ \SI{1}{\joule}), which was necessary due to small execution times for some benchmarks.
In future this will be addressed by balancing the amount of computation required for each benchmark, to standardize the magnitude of results.

All the benchmarks use more energy on the CPU, with the exception of {\tt crc} which as previously mentioned has low floating-point intensity and so is not able to make use of the GPU's greater floating-point capability. 
Variance with respect to energy usage is larger on the CPU, which is consistent with the execution time results.

\section{Conclusions}\label{sec:conclusions}

We have performed essential curation of the OpenDwarfs benchmark suite.
We improved coverage of spectral methods by adding a new Discrete Wavelet Transform benchmark, and replacing the previous inadequate {\tt fft} benchmark.
All benchmarks were enhanced to allow multiple problem sizes; in this paper we report results for four different problem sizes, selected according to the memory hierarchy of CPU systems as motivated by Marjanovi{\'c}'s findings~\cite{marjanovic2016hpc}.
These can now be easily adjusted for next generation accelerator systems using the methodology outlined in Section~\ref{ssec:setting_sizes}.

We ran many of the benchmarks presented in the original OpenDwarfs~\cite{krommydas2016opendwarfs} paper on current hardware.
This was done for two reasons, firstly to investigate the original findings to the state-of-the-art systems and secondly to extend the usefulness of the benchmark suite.
Re-examining the original codes on range of modern hardware showed limitations, such as the fixed problem sizes along with many platform-specific optimizations (such as local work-group size).
In the best case, such optimizations resulted in sub-optimal performance for newer systems (many problem sizes favored the original GPUs on which they were originally run).
In the worst case, they resulted in failures when running on untested platforms or changed execution arguments.

Finally a major contribution of this work was to integrate LibSciBench into the benchmark suite, which adds a high precision timing library and support for statistical analysis and visualization.
This has allowed collection of PAPI, energy and high resolution (sub-microsecond) time measurements at all stages of each application, which has added value to the analysis of OpenCL program flow on each system, for example identifying overheads in kernel construction and buffer enqueuing.
The use of LibSciBench has also increased the reproducibility of timing data for both the current study and on new architectures in the future.

\section{Future Work}\label{sec:future_work}

We plan to complete analysis of the remaining benchmarks in the suite for multiple problem sizes.
In addition to comparing performance between devices, we would also like to develop some notion of `ideal' performance for each combination of benchmark and device, which would guide efforts to improve performance portability.
Additional architectures such as FPGA, DSP and Radeon Open Compute based APUs -- which further breaks down the walls between the CPU and GPU -- will be considered.

Each OpenCL kernel presented in this paper has been inspected using the Architecture Independent Workload Characterization (AIWC).
Analysis using AIWC helps understand how the structure of kernels contributes to the varying runtime characteristics between devices that are presented in this work, and will be published in the future.

Certain configuration parameters for the benchmarks, e.g. local workgroup size, are amenable to auto-tuning.
We plan to integrate auto-tuning into the benchmarking framework to provide confidence that the optimal parameters are used for each combination of code and accelerator.

The original goal of this research was to discover methods for choosing the best device for a particular computational task, for example to support scheduling decisions under time and/or energy constraints.
Until now, we found the available OpenCL benchmark suites were not rich enough to adequately characterize performance across the diverse range of applications and computational devices of interest.
Now that a flexible benchmark suite is in place and results can be generated quickly and reliably on a range of accelerators, we plan to use these benchmarks to evaluate scheduling approaches.

\section*{Acknowledgements}
We thank our colleagues at The University of Bristol's High Performance Computing Research group for the use of ``The Zoo'' Research cluster for experimental evaluation.

\bibliographystyle{ACM-Reference-Format}
\bibliography{bibliography/bibliography}

\newpage
\listoftodos[Notes]

\end{document}